\documentclass[runningheads]{llncs}

\usepackage{url}
\usepackage{float}
\usepackage{graphicx}
\usepackage{amsmath}
\usepackage{amssymb}

\usepackage[utf8]{inputenc}
\usepackage[T1]{fontenc}
\usepackage{textcomp}
\usepackage{xcolor}
\newcommand{\marios}[1]{{\color{black}#1}}
 
\usepackage{tikz}
\usetikzlibrary{graphs,graphs.standard,quotes} 
\usetikzlibrary{decorations.pathreplacing,decorations.markings,positioning}
\tikzset{
 mid arrow/.style={draw, postaction={decorate},
 decoration={
    markings, mark=at position 0.5 with {\arrow{stealth}}}},
 every node/.style={draw,circle}}

\usepackage{graphicx}
\title{Multiple Node Immunisation for Preventing Epidemics on Networks by Exact  Multiobjective Optimisation of Cost and Shield-Value\thanks{Michael Emmerich ackknowledges support by the EU2020 RISE SMA Project.}}
\author{Michael Emmerich\inst{1,2}\orcidID{0000-0002-7342-2090} \and
Joost Nibbeling\inst{1}
\and Marios Kefalas\inst{1} \and
Aske Plaat\inst{1}
}
\authorrunning{Michael Emmerich et al.}
\institute{Leiden Institute of Advanced Computer Science, Leiden University, The Netherlands,
\email{emmerich@liacs.nl}\\
\url{http://liacs.leidenuniv.nl} \and
FINNOPT, Ahlmaninkatu 2
	40100
	Jyväskylä, Finland}
\begin{document}

\maketitle              
\begin{abstract}
The general problem in this paper is vertex (node) subset selection with the goal to contain an infection that spreads in a network. Instead of selecting the single most important node, this paper deals with the problem of  selecting multiple nodes for removal. As compared to previous work on multiple-node selection, the trade-off between cost and benefit is considered. The benefit is measured in terms of increasing the epidemic threshold which is a measure of how difficult it is for an infection to spread in a network. The cost is measured in terms of the number and size of nodes to be removed or controlled. Already in its single-objective instance with a fixed number of $k$ nodes to be removed, the multiple vertex immunisation problems have been proven to be NP-hard. Several heuristics have been developed to approximate the problem. In this work, we compare meta-heuristic techniques with exact methods on the Shield-value, which is a sub-modular proxy for the maximal eigenvalue and used in the current state-of-the-art greedy node-removal strategies. We generalise it to the multi-objective case and replace the greedy algorithm by a quadratic program (QP), which then can be solved with exact QP solvers.
The main contribution of this paper is the insight that, if time permits, exact and problem-specific methods approximation should be used, which are often far better than Pareto front approximations obtained by general meta-heuristics. Based on these, it will be more effective to develop strategies for controlling real-world networks when the goal is to prevent or contain epidemic outbreaks.
This paper is supported by ready to use Python implementation of the optimization methods and datasets.
\end{abstract}

\section{Introduction}
The overarching goal of this paper is to find methods that help to make networks more robust against virus attacks (SIS type epidemics).  This is done by selecting nodes from the network to immunise or to remove such that the largest eigenvalue is reduced and thereby the epidemic threshold is improved \cite{chakrabarti2008epidemic}. In this paper, we derive a quadratic programming version of the problem of reducing the largest eigenvalue of a complex network (represented by the Netshield proxy-function \cite{chen2016node}) and compare the new exact method with heuristic methods proposed previously in \cite{maulana2017immunization}. This strategy is proposed for solving the  bi-objective cost-benefit optimization problem, and we discuss for the first time the exact Pareto front obtained by our new method for different example networks.

The problem that we discuss may for instance arise when combating or preventing the spread of viruses such as recently Ebola or SARS variants \cite{plaat}. To evaluate solutions an appropriate  vulnerability measure is necessary. For this paper the eigenvalue drop (abbreviated with 'eigen-drop') is used which is inversely proportional to the epidemic threshold. More precisely, the \marios{eigenvalue drop}  measures the decrease of \marios{the} maximum eigenvalue of the adjacency matrix
of a network 
and the inverse relationship to the increase of the epidemic threshold \marios{,} under the SIS epidemic model. \cite{chakrabarti2008epidemic} \cite{li2013epidemic}. 
The epidemic threshold is a critical value inherent to the given networks that determines the contagiousness a virus requires to infect the entire network (exponential growth) or to disappear (exponential decay). The SIS model, a is a special case of the so-called contact process where a virus can spread from an infected node to any of its susceptible nod neighbours and infect those neighbours. 
At some later point in time the infected nodes recover and become susceptible to infection again. 

A network or graph $G$ consists a pair $(V,E)$. Here $V$ is a set of nodes $V=(v_1, \dots, v_n)$. $E \subseteq V \times V$ is a set of edges representing connections between the nodes. A graph can also be represented as an adjacency matrix $A(V,E) \in \{0,1\}^{n \times n}$ with $a_{ij} = 1$ if $(v_i, v_j) \in E$ or $a_{ij} = 0$ if $(v_i, v_j) \notin E$. The first or maximum eigenvalue of this graph will be denoted with $\lambda$ and the corresponding eigenvector with $u$. 

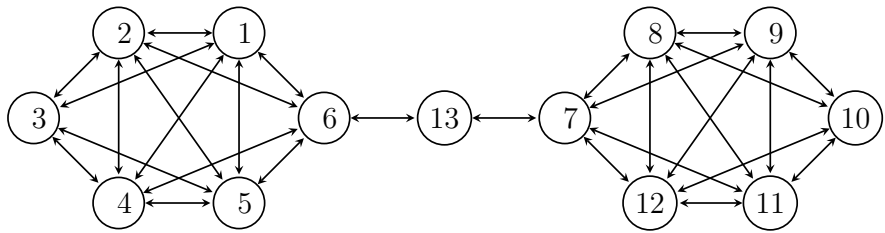
\begin{figure}
\scalebox{0.8}{
\begin{tikzpicture}[<->,>=stealth,shorten >=1pt,auto,node distance=2cm,
                    thick,main node/.style={circle,draw,font=\sffamily\Large\bfseries}]
  \node[main node] (1) {$\ 1$};
  \node[main node] (2) [left of=1] {$\ 2$};
  \node[main node] (3) [below left of=2] {$\ 3$};
  \node[main node] (4) [below right of=3] {$\ 4$};
  \node[main node] (5) [right of=4] {$\ 5$};
  \node[main node] (6) [below right of=1] {$\ 6$};
  \node[main node] (13) [right of=6] {$13$};
  \node[main node] (7) [right of=13]{$\ 7$};
  \node[main node] (8) [above right of=7] {$\ 8$};
  \node[main node] (9) [right of=8] {$\ 9$};
  \node[main node] (10) [below right of=9] {$10$};
  \node[main node] (11) [below left of=10] {$11$};
  \node[main node] (12) [left of=11] {$12$};
\path[every node/.style={font=\sffamily\small}]
  (1) edge [right] (2)
  (1) edge [right] (3)
  (1) edge [right] (4)
  (1) edge [right] (5)
  (1) edge [right] (6)
  (2) edge [right] (3)
  (2) edge [right] (4)
  (2) edge [right] (5)
  (2) edge [right] (6)
  (3) edge [right] (4)
  (3) edge [right] (5)
  (4) edge [right] (5)
  (4) edge [right] (6)
  (5) edge [right] (6)
  
  (7) edge [right] (8)
  (7) edge [right] (9)
  (7) edge [right] (11)
  (7) edge [right] (12)
  (8) edge [right] (9)
  (8) edge [right] (10)
  (8) edge [right] (11)
  (8) edge [right] (12)
  (9) edge [right] (10)
  (9) edge [right] (11)
  (9) edge [right] (12)
  (10) edge [right] (11)
  (10) edge [right] (12)
  (11) edge [right] (12)
  (6) edge [right] (13)
  (13) edge [right] (7)
  ;
\end{tikzpicture}
} 
\caption{\label{fig:example} A small version of the “bar-bell” graph. Two cliques of the same size connected with a bridge. Using this graph, it can be demonstrated that targeted immunization, that is focusing on the removal of high degree nodes, is not always the best strategy in immunization. Adapted from \cite{chakrabarti2008epidemic}.}
\end{figure}

Note, that there that in the literature various strategies for reducing the vulnerability of a complex network to virus attacks have been suggested. Examples are, for instance,  random immunisation, acquaintance immunisation, and target immunisation.
Many of these strategies can be seen as heuristics and do not specifically relate to the spread dynamics of viruses in real world networks.
In Chakrabarti et al. \cite{chakrabarti2008epidemic} it is argued that it is \marios{of} paramount importance to take into account the dynamics of the contact process and to chose an appropriate measure with respect to the global structure of the complex networks.
Local measures, such as the degree of a vertex, do in general lead to mediocre performance and can be even misleading. For example targeted immunisation would choose the nodes with the highest degrees (hubs).  To focus the strategy, at least partially, on lower degree nodes, may seem counter-intuitive.  However, it is not always the case that the immunisation of the highest degree nodes will reduce the vulnerability of the network.
In contrast, it has been shown that focusing on the reduction of the largest eigenvalue of the adjacency matrix of the graph is a effective way to reduce the epidemic threshold which determines whether the number of infections grows exponentially or decreases exponentially.
In particular in the early stages of an outbreak there is a broad consensus of the effectiveness of this strategy.
\marios{In} \cite{chakrabarti2008epidemic} \marios{the authors} provide the example of the barbell graph \marios{(see Fig~\ref{fig:example})} that demonstrates this, where node 13 is of crucial importance although it has low degree. 


\begin{definition}
Given a network $G$ 
and a network $G'$ where $G'$ is a sub-graph of $G$ with some its nodes and adjacent edges removed, $\Delta\lambda$ or eigen-drop 
is defined as the difference between the maximum eigenvalue of the adjacency matrix of $G$ and the maximum eigenvalue of the adjacency matrix of $G'$.
\end{definition}
\begin{definition}
The $k$-node immunisation \marios{problem}: \marios{G}iven a graph $G=(V,E)$ and $k \marios{\in \mathbb{N}}$, \marios{the $k$-node immunisation problem aims in finding} a set of nodes $S \subseteq V$ with $|S| = k$, such that the removal of these nodes from $G$ maximises the eigen-drop $\Delta\lambda$.
\end{definition}

 \begin{figure}[t]
     \centering
     \includegraphics[width=\textwidth]{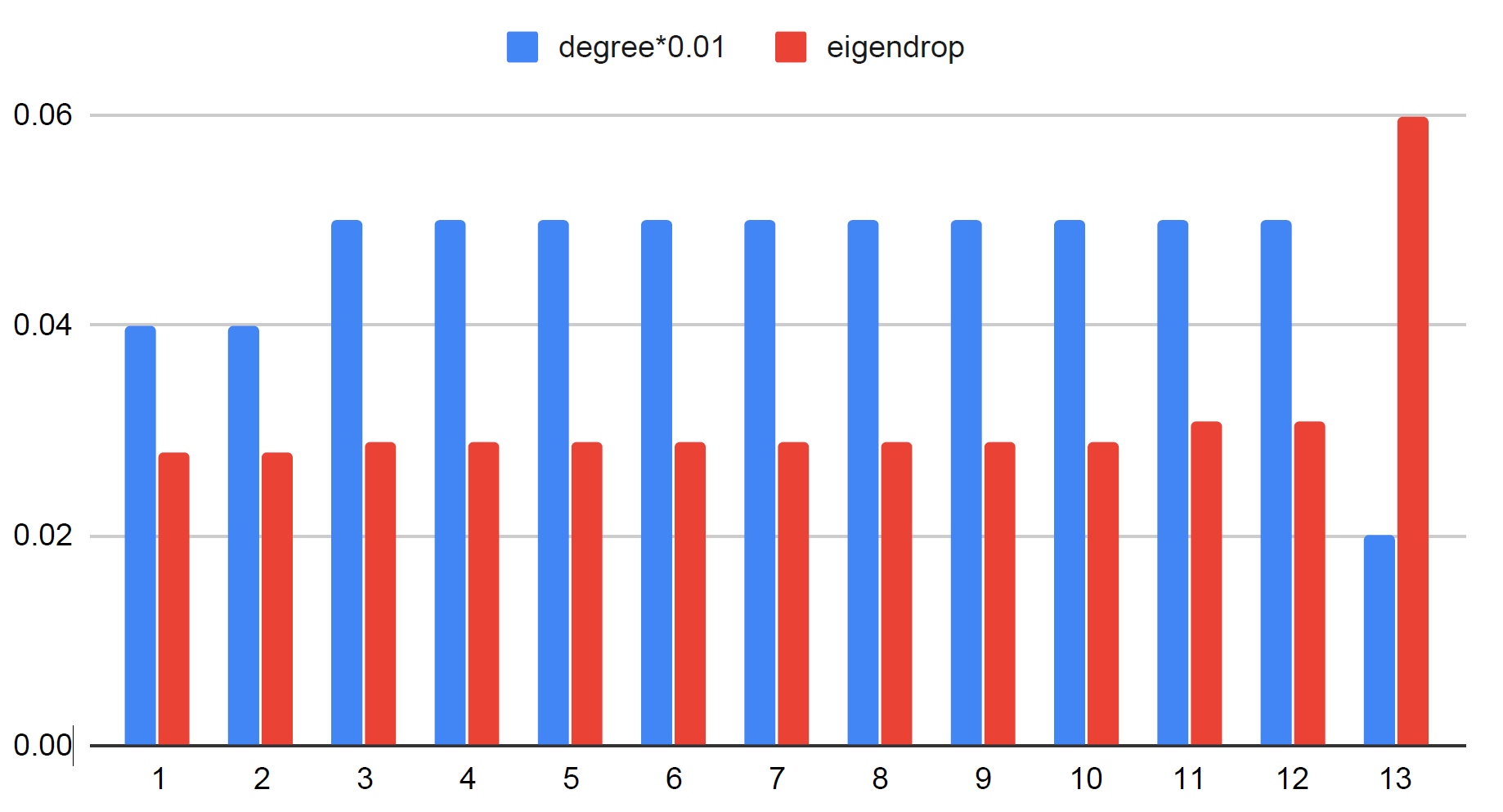}
     \caption{Table with eigen-drops vs. degrees of extended barbell graph}
     \label{fig:eigendroptable}
 \end{figure}
It has been shown in \cite{chen2016node} that this problem is NP-hard. Therefore heuristic methods have been suggested for solving this problem such as the NetShield and NetShield+ algorithms in \cite{chen2016node} and a problem specific genetic algorithm in \cite{maulana2017immunization}. 
The drawback of these heuristics is that they are designed for the $k$-node immunisation problem which requires that a good value of $k$ is known in advance.
In addition, the heuristics treat\marios{this problem} as if every node requires the same effort to remove. Therefore in this paper, we reformulate the $k$-node immunisation problem to a multi-objective one.
This is similar to [7], where multi-objective optimisation using an evolutionary optimisation heuristic was used to take into account the cost of removal. 
Indeed, the nodes with maximal eigen-drop do not coincide with the nodes with a high degree, as is seen in Figure \ref{fig:eigendroptable}. Node 13 in the earlier discussed barbell graph has only a degree of $2$ but it significantly reduces the Eigenvalue.
\begin{definition}
The multi-objective immunisation problem given a graph $G=(V,E)$, a cost denoted with $\mathrm{Cost}(v)$
for each $v \in V$ and $S \subseteq V$ reads
\begin{eqnarray}
    f_1(S) = \Delta\lambda &\rightarrow& \max\\
    f_2(S) = \sum_{v \in S} \mathrm{Cost}(v) &\rightarrow& \min
\end{eqnarray}
\end{definition}
This formulation requires no value for $k$ to be known a-priori and takes the cost of removal in account. As this is a multi-objective problem we are now interested in finding the efficient set and its corresponding Pareto front. To approximate this Pareto front we use and evaluate four different methods. The first two extend the NetShield and NetShield+ algorithms by substituting the first objective with the heuristic used by these methods and then applying the $\epsilon$-constraint method. The second two are two different genetic algorithms specifically designed for multi-objective optimisation problems.

\section{Multiobjective Quadratic Programming Formulation}

The first method for approximating the Pareto front of the multi-objective immunisation problem is based on the NetShield and NetShield+ algorithms designed by Chen Chen et al. \cite{chen2016node}. These methods are designed for the $k$-node immunisation problem and are briefly discussed here. Core to the design of these algorithms is a function called Shield-value.

\begin{definition}
Given the adjacency matrix $A$ of a graph $G(V,E)$, its first eigenvalue $\lambda$, the corresponding eigenvector $u$ and an input set of nodes $S \subseteq V$, the Shield-value function is defined as:
\begin{equation}
    Sv(S) = \sum_{i \in S} 2 \lambda u_{i}^{2} - \sum_{i,j \in S} 2 u_{i}u_{j}A_{ij}
\end{equation}
\end{definition}

The Shield-value function gives an approximation of $\Delta\lambda$ if all nodes in the set $S$ were to be removed from the graph. The NetShield algorithm then finds a set of $k$ nodes that approximates the maximization of this function via greedy selection.

As the cardinality of $S$ grows, the Shield-value function becomes less accurate. The NetShield+ algorithm therefore introduces an extra parameter called the batch size. Instead of finding $k$ nodes at once, a set of $b$ nodes is found and added to the solution. Then these nodes are removed from the network. The new network is used to compute a new Shield-value function, that is again maximized for a set of $b$ nodes. These are then again added to the solution set and this process continues until $k$ nodes have been removed from the network.

To extend these algorithms to work with the multi-objective immunisation problem, we substitute the eigen-drop objective with the Shield-value function. Furthermore, we define the problem as a quadratic multi-objective program by representing the solution with a binary vector $x$. If the node $i$ is in the solution set, $x_i$ will be 1. Otherwise $x_i$ will be 0:

\begin{definition}
Given the adjacency matrix $A$ of a graph $G(V,E)$, its first eigenvalue $\lambda$, and the corresponding eigenvector $u$, the Shield-value function with cost objective is:
\begin{equation}
    f_{1}(x) = \sum_{i=1}^{m} 2 \lambda  u_{i}^{2} x_{i} - \sum_{i=1}^{m} \sum_{j=i+1}^{m} 2 u_{i} u_{j} A_{ij} x_{i} x_{j} \to \max \\
\end{equation}
\begin{equation}
    f_{2}(x) = \sum_{i=1}^{m} x_{i} \mathrm{Cost(i)} \to \min\\
\end{equation}
Subject to:
\begin{equation}
x \in \{0,1\}^{m}
\end{equation}
\end{definition}
To approximate the Pareto front the $\epsilon$-constraint method can be applied \cite{KaisaNMO}. For this method, one of the objectives is transformed into a constraint smaller or equal than $\epsilon$. In this case it will be the second cost objective.
\begin{definition}
Given the adjacency matrix $A$ of a graph $G(V,E)$, its first eigenvalue $\lambda$, the corresponding eigenvector $u$, and some value of $\epsilon$, the Shield-value function with cost constraint is:
\begin{equation}
    f_{1}(x) = \sum_{i=1}^{m} 2 \lambda  u_{i}^{2} x_{i} - \sum_{i=1}^{m} \sum_{j=i+1}^{m} 2 u_{i} u_{j} A_{ij} x_{i} x_{j} \to \max \\
\end{equation}
Subject to:
\begin{equation}
    f_{2}(x) = \sum_{i=1}^{m} x_{i}  \mathrm{Cost}(i) \leq \epsilon \\
\end{equation}
\begin{equation}
    x \in \{0,1\}^{m}
\end{equation}
\end{definition}

By choosing a concrete value of $\epsilon$, the problem is transformed into a quadratic program with a linear constraint. Problems such as these can be solved with a quadratic problem solver via branch-and-bound based methods. By solving multiple programs with different values of $\epsilon$, the Pareto front of the multi-objective Shield-value problem can be found. As the Shield-value is an approximation of $\Delta\lambda$, this Pareto front should therefore also be an approximation of the original problem.

This method can be extended analogously to how the original NetShield algorithm can be extended to NetShield+. Instead of finding a set of nodes that maximises the Shield-value objective at once, an extra batch size parameter $b$ can be introduced. Then a solution that maximises the Shield-value function with only $b$ nodes can be found. These nodes are then added to the complete solution and removed from the network. A new quadratic program can be created with a new Shield-value function computed from the new network. This process continues, adding $b$ nodes to the solution set at every step. This process stops when no more nodes can be added. This occurs when either all nodes have already been added to the solution set, or if any of the nodes not yet added would violate the cost constraint.

\section{Genetic Algorithms}
The advantage of using genetic algorithms over the NetShield based methods described in the previous section, is that they can be made to work directly on the eigen-drop. This sidesteps the need of using a possible inaccurate approximation of the eigen-drop. In addition, by sampling the search space in an efficient manner, genetic algorithms can also consider more candidate solutions that the NetShield methods will. Therefore, it is possible that better Pareto front approximations can be found by these meta-heuristics. 
The GAs used in this paper are specifically designed for multi-objective problems. They use specialised selection operators that aim for both convergence to the Pareto front and spread over the Pareto front. The GAs used are NSGA-II \cite{deb2002fast} and SMS-EMOA \cite{emmerich2005emo}.
In addition to this, it is also possible to hybridise the GAs with the NetShield methods. This is done by initialising the GAs with the solutions found by the NetShield methods. This can cut out the potentially large search effort by the GAs to converge on the Pareto front by starting them from what already is a good approximation. Then the GAs may further refine the solutions using their advantages over the NetShield methods.

\section{Experiments and Results}

A specific cost function is required to define $f_2$.
This cost function should be a good local measure for the effort required for the removal of a node. The cost function we used is the degree of each node, as a highly connected node is likely to be more difficult to remove from the network than a node with less incoming and outgoing edges.

All of the GAs were run 5 times under each configuration, both when the population was initialised at random and when it was initialised with the NetShield solutions. The populations were set to a size of $100$ for the random initialisation. The mutation probability $p_m$ was set to $1/n$, with $n$ being the number of vertices of the graph. Crossover probability $p_c$ was set to $0.75$. All GAs were run with $10000$ iterations of the main loop. 
All results of the GAs are plotted as the first attainment curve \cite{Fonseca96a}. All points on these curves are weakly dominated by only 1 run out of the 5 and are therefore a best case scenario of the GAs

Both the NetShield and NetShield+ methods with the $\epsilon$-constraint method were tested. For the NetShield+ method, the batch size was set to 1. The resolution of the Pareto front approximation depends on how many different values of $\epsilon$ are sampled. As the cost function chosen uses only non-negative integers, it is possible to get the best possible resolution by sampling only a finite amount of points: from 0 to the sum of all degrees increasing $\epsilon$ by 1 every step. The quadratic program solver used is Gurobi\cite{bixby2011gurobi}.

\begin{figure}[t]
\includegraphics[width=\textwidth]{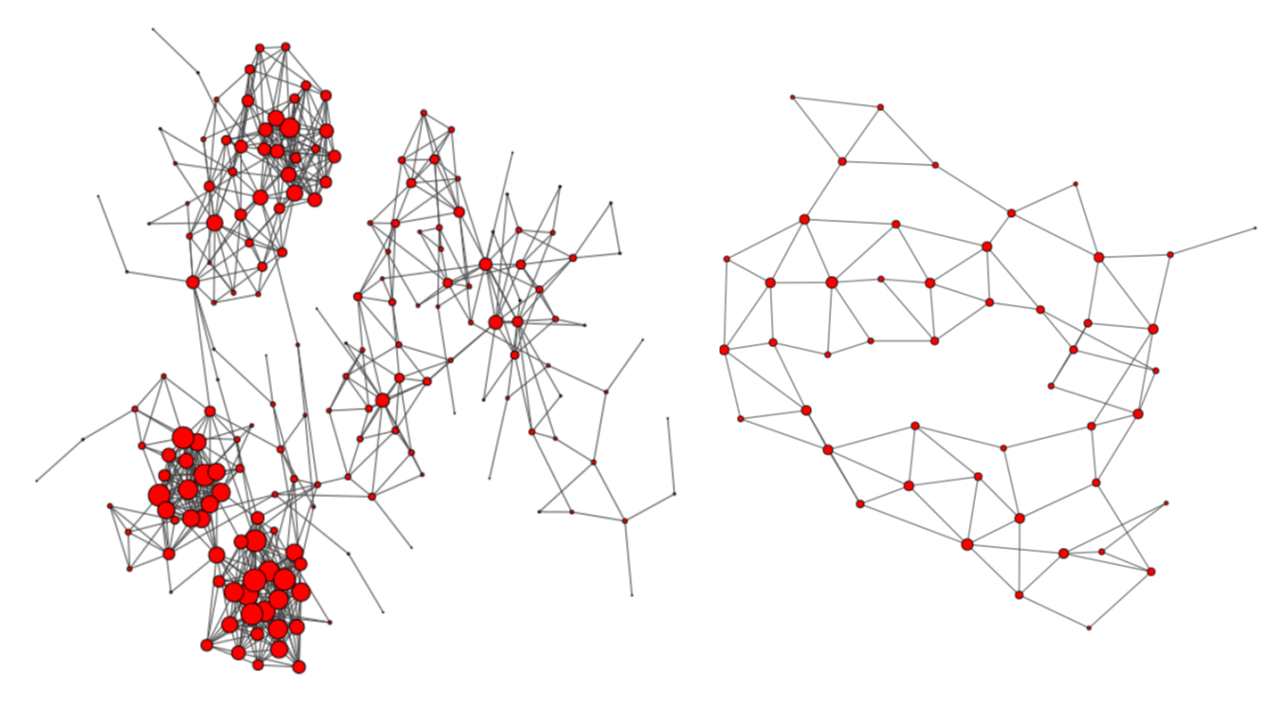}
\caption{\label{fig:conf1pand}Conference Day 1 and Pandemic Network.}
\end{figure}

All results shown are for the following set of four graphs:

\begin{enumerate}
    \item  \textbf{Pandemic}: Based on the Pandemic board game in which a global virus outbreak is fought. The graph connects 27 cities in the world to each other with 93 edges. \cite{maulana2017immunization}. See Figure \ref{fig:conf1pand} (right).
     \item \textbf{Conference Day 1}: Interactions between members of a conference on the first day. Only the largest connected components has been selected from this graph. The graph consists of 190 nodes and 703 edges. See Figure \ref{fig:conf1pand} (left). Taken from \url{www.sociopatterns.org/datasets/infectioussociopatterns}.
     \item \textbf{Erd\H{o}s-R\'enyi graph}: Graph sampled from the Erd\H{o}s-R\'enyi random graph model. This graph has 100 nodes and 294 edges.
     \item \textbf{Barab\'asi-Albert graph}: Graph sampled from the Barab\'asi-Albert graph model.  This graph also has 100 nodes and 294 edges. See Fig. \ref{fig:bara_j1}
\end{enumerate}
\subsection{NetShield with $\epsilon$-constraints and GAs}
\begin{figure}
  \centering
    \includegraphics[width=\textwidth]{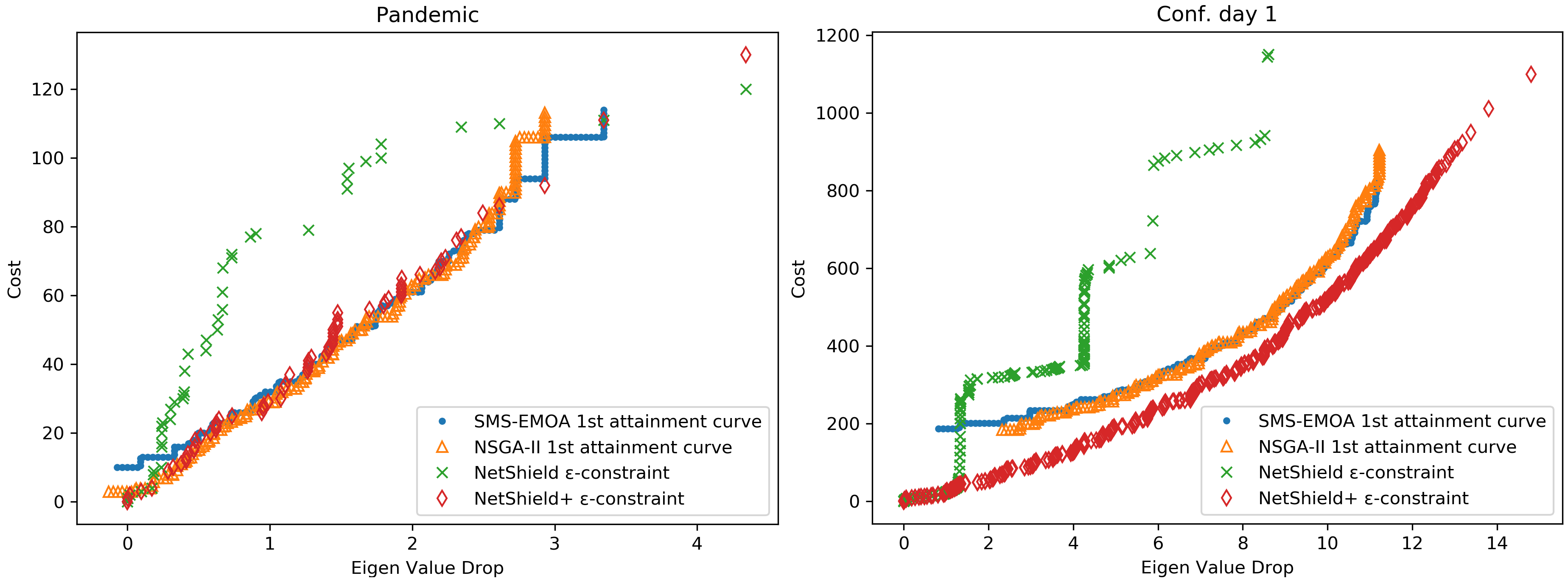}
  \caption{Results GAs and NetShield(+) with $\epsilon$-constraint method}
  \label{fig:res_panconf}
\end{figure}

\begin{figure}
  \centering
    \includegraphics[width=\textwidth]{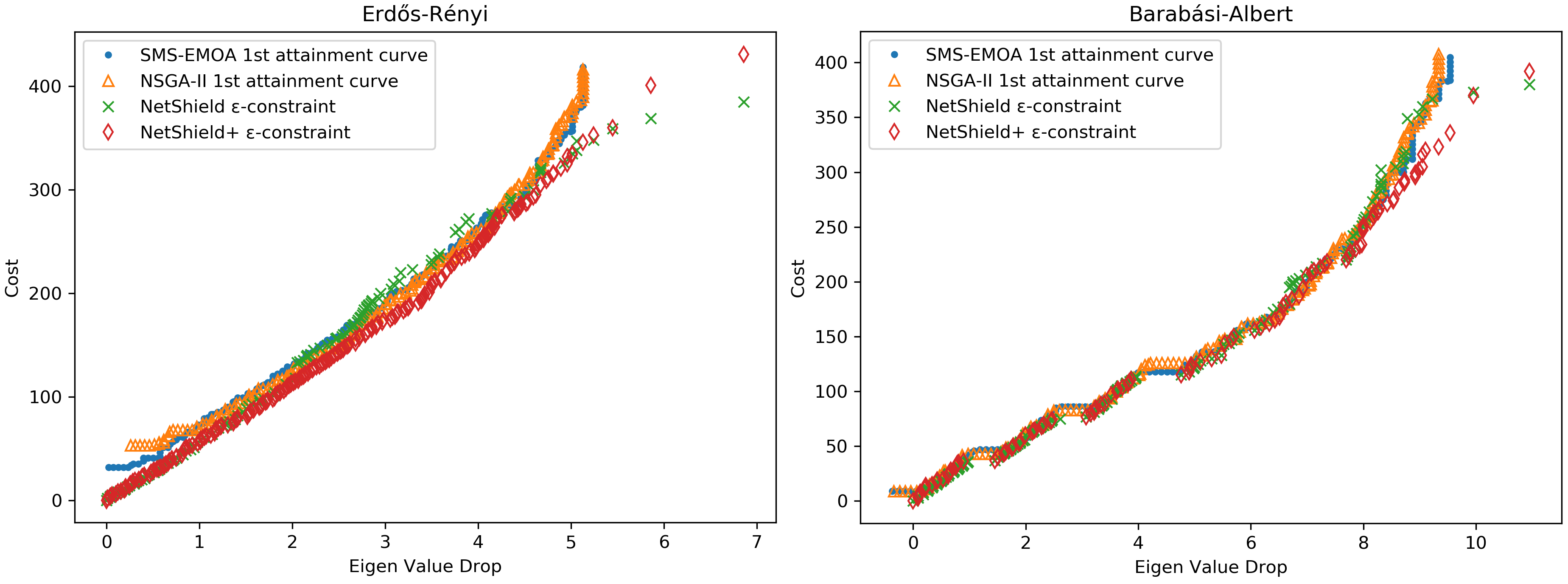}
  \caption{Results GAs and NetShield(+) with $\epsilon$-constraint method}
  \label{fig:res_erbara}
\end{figure}

The results of both the NetShield methods and the randomly initialised GAs are plotted in Figure \ref{fig:res_panconf} for the Pandemic and Conference day graph and in Figure \ref{fig:res_erbara} for the Erd\H{o}s-R\'enyi and Barab\'asi-Albert graph. When comparing the NSGA-II algorithm to the SMS-EMOA algorithm, no clear differences show. It changes from graph to graph which algorithm finds the better Pareto front approximation and they consistently lie very closely together.

Difference do show when comparing the NetShield with the NetShield+ method. Sometimes the difference are large, such as for the Conference day 1 graph and Pandemic graph. For the Barab\'asi-Albert and Erd\H{o}s-R\'enyi graphs the differences are smaller, but the NetShield+ method still tends to give the better results. This is likely due to the Shield-Value losing accuracy when the number of nodes removed increases. Initially the performance of NetShield is very similar to NetShield+. When the allowed cost increases and consequently more nodes can be selected, the NetShield+ method can find solutions that are significantly better. 

At the rightmost extremes however, the NetShield method sometimes finds some solutions that dominate those found by the NetShield+ method. See, for example, the Erd\H{o}s-R\'enyi graph and the Barab\'asi-Albert graph. This may be because the NetShield+ method with a batch size of 1 is more greedy than the NetShield method. With a batch size of 1, the NetShield+ method selects at every step the node with the highest eigenscore that would not violate the $\epsilon$-constraint when added to the solution. It then recomputes a new Shield-value function with the node removed. The NetShield method however, only computes the Shield-value function at the beginning and selects multiple nodes at once to optimise this function. In this way it can take a more global view of the problem. If the Shield-Value then happens to still be a good approximation of the 
eigen-drop, better solution may be found. A possible approach is to repeat the NetShield+ method several times with different batch sizes if time allows. Then all results can be combined for the most accurate Pareto front approximation.

When comparing both the NetShield methods with the GAs, the GAs give very competitive performance when the networks have relatively few nodes. This means the search space is smaller and the GAs have enough time to converge on the Pareto front. This results in some parts of the Pareto front being approximated better by the GAs, because they can work directly on the eigen-drop. This is most notably the case for the Pandemic graph. Here both NetShield method and the NetShield+ method to a lesser degree have difficulty approximating the Pareto front. This is likely caused by this graph having a low maximum degree. This means that the Shield-value approximation loses accuracy quickly \cite{chen2016node}.

The results also show that the Pareto fronts do not form a single distinctive shape. The Pareto front for the Erd\H{o}s-R\'enyi graph is mostly linear. At the rightmost extreme however, two solutions are found where large gains in eigen-drop can be made with a comparatively small cost increase. This results in the Pareto front having a concave section at the end. This is the opposite for the Pareto front for the Conference day 1 graph. For this graph there is a clear case of diminishing returns: it costs increasingly more to get the same improvement in terms of eigen-drop the further the eigen-drop increases. 

\begin{figure}[t]
  \centering
    \includegraphics[width=\textwidth]{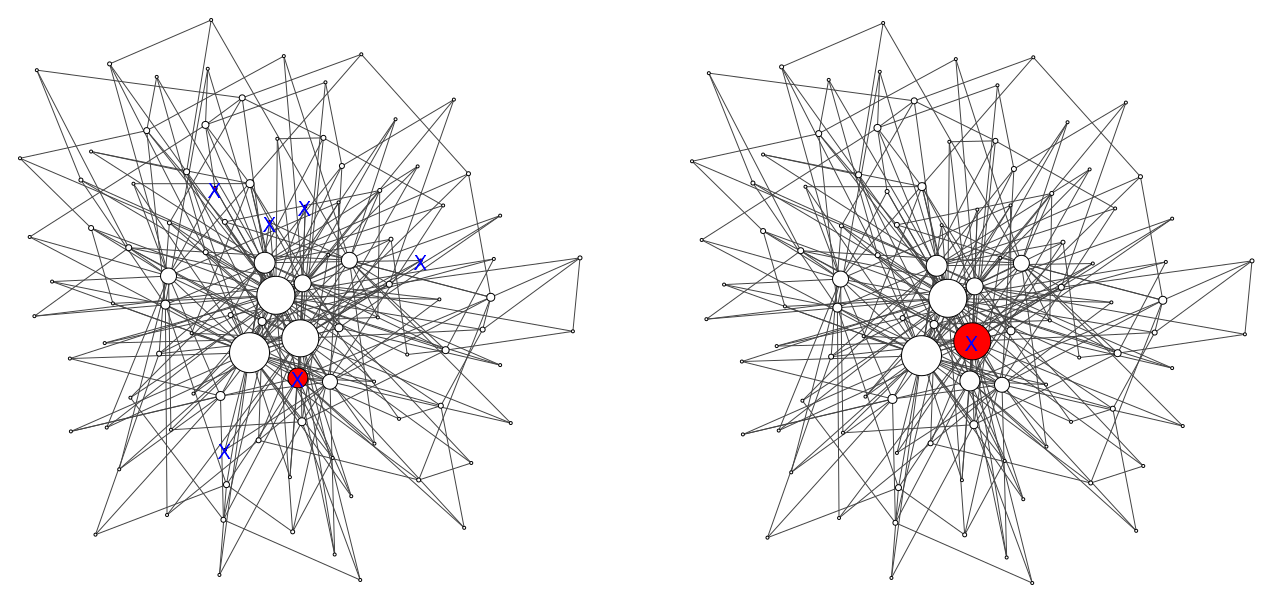}
    \caption{Selected nodes are red and denoted with $\times$. Nodes have been scaled with degree. Left: 6 selected nodes, cost of 36, $\Delta\lambda$ of 0.975.
            Right: 1 selected node, cost of 37, $\Delta\lambda$ of 1.455}
  \label{fig:bara_j1}
\end{figure}

The Barab\'asi-Albert graph Pareto front consists of several sections that are mostly linear, but with gaps in between this sections. At these gaps, large increases in eigen-drop are suddenly gained for low costs. This is the result of the preferential attachment model used to generate this graph. At those points the value of $\epsilon$ allows replacing a larger selection of smaller cost nodes with one of the highly connected hub nodes. While these nodes have high cost, their impact on eigen-drop is still disproportionate to their cost. Two solutions with this graph are visualised in figure \ref{fig:bara_j1}: one at the left side of a gap and one at the right side.

\subsection{Hybrid GA approach}
\begin{figure}
  \centering
    \includegraphics[width=\textwidth]{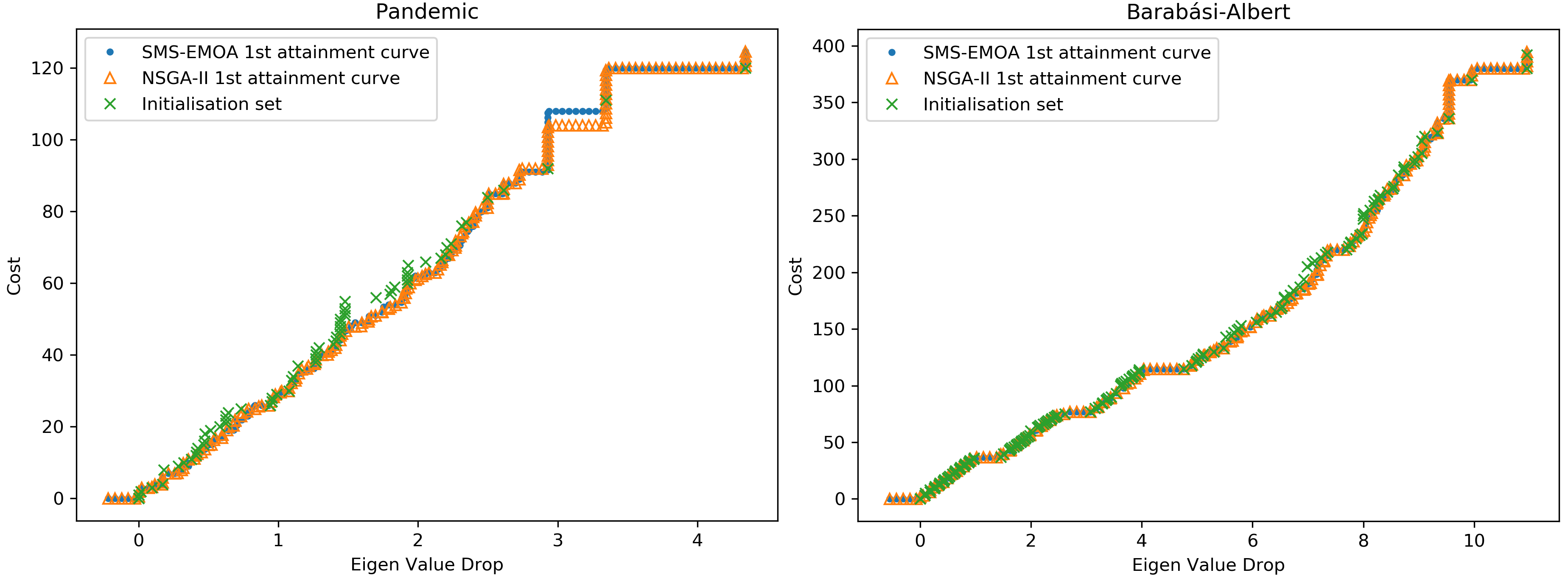}
  \caption{Results hybrid GAs}
  \label{fig:res_atins}
\end{figure}

The results with the GAs initialised from the results from the NetShield methods for the Pandemic and Barab\'asi-Albert graph are shown in Figure \ref{fig:res_atins}. They are shown together with the initialisation sets. The most notable improvements found by the GAs are for the Pandemic graph. Here the results of the inaccuracies of the Shield-value have been corrected. It appears that in these cases, the GAs have the ability to repair such issues.
The improvements for the Barab\'asi-Albert graph are more minor. Either the initialisation sets are already close to the Pareto fronts or there may not be enough diversity in the initial populations for the GAs to find better solutions.
\section{Conclusion}
In this paper, it is shown that the NetShield and NetShield+ algorithms can be extended with an $\epsilon$-constraint method, using an exact quadratic programming solver (here: Gurobi). In this manner, a multi-objective variant of the node immunisation problem can be solved with a cost function added which is proportional to the effort of the node removal. The performance is mostly equivalent or better than two multi-objective genetic algorithms specifically designed for multi-objective optimisation, except for cases where the Shield-value function is inaccurate due to characteristics of the network. In general the NetShield+ method is more robust than the NetShield method, but there are exceptions. Combining the GAs with the NetShield algorithm as initial population only provided small further improvements. Therefore, if time permits and the most accurate Pareto front approximation is required, a valid approach would be to use all methods and combine the end results.
The results also show that there does not appear to be a typical shape to the Pareto fronts resulting from this problem. They are dependent on both the topology of the network and on the cost function.

A main contribution of this paper is the insight that, if time permits, exact and problem-specific methods approximation should be used, which are often far better than Pareto front approximations obtained by general meta-heuristics. 
Based on these insight, it will be more effective to develop strategies for controlling real-world networks when the goal is to prevent epidemic outbreaks.
It should be noted, however, that we focused solely on the eigen-value drop. While this is an effective measure under the SIS infection model, the properties of real world epidemics may not be fully captured by this model. Such additional aspects are mentioned, for instance, in \cite{plaat}. In addition, we also assume that any nodes in the network can be immunised.  This also may not translate well to real world scenarios. Therefore, future work should also take a broader view of the problem than further improving the eigen-drop via the process of node removal.

A note of precaution shall be provided here, when applying the model to epidemics in a late stage. A crucial assumption is that the epidemic is in an earlier stage, which will imply that it is likely many neighboring nodes of an infected node are not yet infected. In a late stage of an epidemic this can no-longer be assumed. Moreover, the dynamics are \marios{of the} susceptible-infected-suceptible (SIS) type, and not of the susceptible-infected-recovered (SIR) type, which would be also a common real-world model, e.g., because it would be applicable to the recent SARS-CoV2 pandemic. 
It is conjectured, that also in the SIR dynamics the largest eigenvalue is an important value to focus on when dampening or preventing a virus outbreak. When it comes to more realistic and fine grained models of network dynamics, it is probably not straightforward anymore to use a single indicator, such as the eigen-drop, and more complex simulation models are required such as the event-based simulation model in \cite{michalak2020evolutionary}. For an excellent overview various  epidemic models and dynamics on complex networks the reader is referred to \cite{pastor2015epidemic}.

Whereas in our study exact QP solvers were applicable due to the quadratic equations in the objective function, in the more general setting this will no longer be the case and black-box optimization, such as SMS-EMOA will be very useful. 
\paragraph{Software*}: All source code (Python) of algorithm implementations and network data of this study is made free available under:\\
\url{https://github.com/joostnibbeling/node-immunisation}    

\addcontentsline{toc}{section}{References}

\bibliographystyle{abbrv}
\bibliography{bibliography}

\end{document}